# Can LLMs Produce Original Astronomy Research in a Semester? A Graduate Class Experiment

Ann Zabludoff[1], Chen-Yu Chuang, Parker Thomas Johnson, Yichen Liu, Brina Bianca Martinez, Neev Shah, Lucille Steffes, Gabriel Glen Weible

Department of Astronomy and Steward Observatory[2], University of Arizona

## Summary

We discuss the results of using large language models (LLMs) to conduct original scientific research in an unfamiliar subject area during the Fall 2025 semester. Students in a graduate astronomy and astrophysics course were asked to test whether LLMs could help them complete research tasks faster and at a level of detail and accuracy required for scientific publication. Most students employed LLMs for a total of 5–10 hours. While all students completed a draft paper on an unsolved problem related to galaxies by semester's end, their impressions of the models' value varied. About half thought that the models saved them time. Many noted that LLMs failed to provide appropriately detailed insights or steps to addressing open, niche questions over a several-month timeframe. The LLMs also frequently (about 20% of the time) returned false citations, links, or summaries of papers. The models struggled with generating complex functional code, accessing online packages or Application Programming Interfaces (APIs), and retrieving astronomical datasets from existing archives. In writing code and in chats, the LLMs made implicit, overly simplifying assumptions and often doubled down even after being corrected. Given the rapid pace of LLM development, new models may soon address at least some of these issues and thus significantly enhance research productivity. Yet students expressed concerns about how LLM use might dampen creativity and reflection during the research process. To improve learning experiences in future semesters, the class will first discuss LLM best practices and limitations. Students will be encouraged to explore free online resources for tips for generative model applications and will decide for themselves whether to use LLMs for their research project.

This white paper was not written using LLMs.

## Motivation

Current LLMs are known to exclude or misattribute scientific papers and results, as well as return fake data and often subtly inaccurate code. The models also lack scientific "taste," i.e., the judgement or intuition—developed by scientists after years of work and literature study—to choose the most interesting and tractable problems. Yet LLMs can be remarkably helpful for learning scientific concepts, synthesizing the domain literature, and quickly ascertaining whether a research project is potentially new and viable. We conducted an experiment in Fall 2025 for Astronomy and Astrophysics PhD students in a graduate Galaxies course: using LLMs to develop and complete individual research projects by the end of the semester. The aspirational goal was not a term paper summarizing past work, but a draft manuscript containing sufficiently novel and accurate results to justify submission to a peer-reviewed journal.

## Background

Students from many nations and undergraduate institutions attend the University of Arizona graduate Astronomy and Astrophysics program, typically obtaining a PhD in 5–6 years. All have a strong undergraduate preparation in physics and mathematics, and some have masters degrees in these or related fields from other universities. The coursework during the first two years of graduate study is rigorous and designed to prepare students to conduct research across astronomy and astrophysics.

---

[1] Corresponding author: aiz@arizona.edu
[2] 933 N Cherry Ave, Tucson, Arizona, 85721 USA

ASTR 540 Structure and Dynamics of Galaxies is a first- or second-year graduate course that draws material from *Galaxies in the Universe: An Introduction* by L.S. Sparke and J.S. Gallagher, *Galactic Astronomy* by Binney and Merrifield, and *Galactic Dynamics* by Binney and Tremaine. The lectures introduce results from relevant papers from the literature that are accessible via the arXiv[3] and NASA Astrophysics Data Server (ADS)[4]. ASTR 540 touches broadly on many topics in galactic and extragalactic astronomy, including:

- Structure and kinematics of the Milky Way
- Resolved and unresolved stellar populations
- Spiral galaxy structure, kinematics, gas and stellar populations
- Elliptical galaxy morphology, kinematics and stellar populations
- Stellar populations and chemical evolution in the Milky Way
- Galaxy dynamics including orbits, potentials, relaxation
- Interstellar medium structure and content, i.e., HI, $H_2$, ionized gas, particles, magnetic fields
- Black holes and active galaxies
- Stellar populations in the Milky Way
- Chemical evolution: observations and models
- Clues for galaxy evolution from the Local Group
- High-redshift galaxies: identification and properties
- Galaxy formation and evolution, star formation history, and star formation rate

The suggested use of LLMs in the syllabus read as follows: "During the term, on your own or with the help of a LLM, you will identify a potentially publishable research project on galaxies. You will use your own search to confirm that similar work has not yet been done. You will then employ the LLM to help you grab and plot the relevant data or perform the appropriate calculation. You will check that the results are accurate and complete. Finally, you will draft a paper on Overleaf in your own words and give a five-minute presentation on the results. Your presentation will have a maximum of three slides and will: 1) motivate the work, i.e., why it is interesting and better than what was done before, 2) include a key figure or two, and 3) explain your result and the checks you conducted."

**Methodology**

Here we discuss examples of how the students identified a research topic and obtained new results.

At the start of the course in late August 2025, all seven students were in their first year of graduate study. Most (5/7, ~70%) were pursuing PhD-track research in branches of astronomy other than galaxy formation, evolution, structure, or dynamics. Their previous experience with LLMs varied from limited to moderate. Students had used LLMs as a supplemental study aid and for research-related coding and debugging, e.g., to generate plots or data analysis scripts. All had access to OpenAI ChatGPT-4o, 5, or 5.2, Anthropic Claude Sonnet 4 or 4.5, Opus 4, or Haiku 4.5, and/or Google Gemini Flash or Pro 2.5 or 3.

For their ASTRO 540 individual research projects, students described using LLMs for each stage of the project and typically for a total of 5–10 hours, including time spent exploring what the generative models could and could not do well. Several students used more than one model, e.g., ChatGPT-5 + Claude Sonnet-4 or ChatGPT-5.2 + Claude Opus-4 + Gemini Pro-3. One student who employed three models totalled about 30 hours of usage and noted that the LLMs took longer to think as the project grew, from a few seconds to tens of minutes of thinking per prompt.

The individual research projects developed by the students ranged broadly (see Appendix). Students generally used the LLMs to select and refine their research topic and to design the experiment and analysis pathway. In particular, students reported using the models for synthesizing relevant literature, including finding the open questions in the field, brainstorming research questions and the methodologies for each research task, identifying relevant datasets and scientific models, and assisting with writing code for simulations or analyses. Besides creating code, LLMs provided insight into how to use existing codes and to extract specific outputs. One student broke their LLM usages into two general categories: 1) code-level help, i.e., explaining what scripts do, rewriting/refactoring Python, and generating reproducible snippets for astronomy pipelines, and 2) "agentic"/multi-step workflows for research engineering, specifically LLM-driven code-generation loops (e.g., a Gemini-API orchestrator with reviewer/feedback iterations) to help build and debug components of their analysis/simulation tooling.

---

[3] https://arxiv.org/archive/astro-ph
[4] https://scixplorer.org

Students verified the model results in several ways, including manually checking every literature source with Google or NASA ADS, reading the relevant sections of all the sources identified with the LLMs in order to cite their material directly, not from a generated summary. Students also went back to primary sources, compared the proposed methodologies to what was in the literature, and tested that the model-generated code matched the expected outputs. As a check on model-generated literature citations, one student read through each paper abstract returned by the LLM and asked the LLM for the section of the paper where the information originated, making it easier to confirm that the LLM returned what was actually in the paper. As a code check, another student let the LLM explain what it was doing for every few agentic iterations and then checked for errors in the explanation. The student also employed different LLMs to cross-check each other. When an LLM could not resolve a code issue, the student manually reviewed and repaired the code.

**Successes**

Students found the LLMs broadly useful. The LLMs allowed them to efficiently synthesize relevant literature, identify key papers, and generate research questions and methodologies, which significantly sped up those initial stages of the project. One student commented that the model allowed them to narrow down the specific goal of their project and to find tools that could help answer their research question. Another student found it helpful that the LLM suggested potential next steps in what the model had been prompted to do. Students stressed the usefulness of LLMs for writing code for small and specific tasks, such as producing images related to their research, and for checking for simple code bugs. The LLMs also helped draft sections of their papers, improving writing efficiency.

Students experienced several positive surprises when working with the LLMs, including the transformation of what would have been a days- or weeks-long literature search into an instant synthesis of complex information from various sources. The summaries and insights provided by the models helped the students to decide on novel research topics and guided their research direction. Students expressed surprise that the LLMs seemed well-trained on astronomy and astrophysics literature, noting that the models' wide, deep knowledge and ability to answer basic questions exceeded the students' own background at the start of the project. While students were generally familiar with the models' ability to develop and explain code, one student was impressed that ChatGPT found code errors just from ingesting a scientific plot of their data. In that case, uploading the energy and angular momentum evolution plot derived for a triple star system led the model to recognize unphysical changes and to point out that the secular equation in the code was wrong.

**Failures**

The failures of the LLMs fell into several general categories. The first type of failure was that the models struggled with generating complex functional code for data analysis and physically accurate code for simulations, requiring extensive manual correction. When generating code for a student's dynamical simulation, Claude Opus 4 was more successful than Gemini Pro 3 and the ChatGPT-5 thinking model—it appeared to better predict how each code component contributed to the final outcome, which allowed it to diagnose and fix problems that the other LLMs missed. However, Claude's lower usage limit prevented employing it as much as necessary.

The student attempted to introduce a flyby of a dark matter subhalo into the simulation code, but frequently encountered non-physical results. ChatGPT and Gemini were unable to diagnose the problem, which the student ultimately traced to an incorrect implementation of the Milky Way's tidal field, as well as to an incompatible formatting of the subhalo density profile within the simulation package. For the tidal field, the LLMs had mixed parameters from two different papers and generated code that produced an unphysical potential. The student ultimately resolved the issues, correcting the tidal setup by replacing the parameters to make them consistent and implementing the subhalo profile in a package-compatible format. The student concluded that the LLMs struggled more to debug code when no documentation was available and that the models did not necessarily examine the entire codebase before generating solutions, leading to errors or incomplete fixes.

The second failure mode was that the LLMs frequently–about 20% of the time—created links to the wrong sources or misattributed papers, requiring extensive manual review. In a significant number of cases, even when the URL was accurately generated by the LLM, the model only returned the correct person and year, and the title and summary were different from the actual paper. For example, in one instance, Claude Sonnet 4.5 was prompted "What works have studied relationships between YSO [Young Stellar Object] properties (e.g., demographics) as a function of gas-phase metallicity in the MW [Milky Way] and/or the Local Group galaxies?." The LLM first returned an apparent 2025 James Webb Space Telescope study examining YSOs in the open star cluster NGC 346 in the Small Magellanic Cloud (SMC). However, the two links that Claude provided for this reference both directed the student to an online section of "Outskirts of Galaxies"[5] on the chemical abundances of H II regions in outer galaxy disks. The linked webpage was unrelated to how Claude presented it, lacking any mention of YSOs, NGC 346, or the SMC. The student found other similarly mislinked references in the same response from Claude.

A third issue was that even when the models did point, say, to a correct link to data tables, they did not successfully query the tables themselves. Despite agentic advances, this apparent inability to directly search the internet or to interact with APIs, e.g., for querying data tables and downloading data from them, is a significant impediment to research. The LLMs did not demonstrate familiarity with the technical details of such datasets and how they were organized internally, beyond being able to point to a URL. For example, the models had difficulties autonomously accessing and retrieving astronomical datasets from online archives like VizieR (Ochsenbein+99) and running online software packages such as STARBURST99 (Hawcroft+2025). In another example, the LLMs generated code that was incompatible with the MaNGA (Bundy+2015) dataset format.

A fourth problem was that the LLMs made implicit, overly simplifying assumptions and often doubled down even after a student pointed out inaccuracies. For example, a student asked for different theoretical initial stellar mass functions to be plotted on the same graph with links to papers in the literature. Instead, the LLM plotted a single and basic initial mass function without any references, explaining only that its answer fit the general shape of an initial mass function. In another case, an LLM kept pointing towards lines in a piece of code that did not exist. When asked to determine which databases hosted the archival data that a student needed and how to access it, an LLM repeatedly identified the wrong database and confidently wrote code to access data from non-existent archives. When a student searched for data about the metallicities of molecular cloud regions, the LLM provided values without references or errorbars, even after multiple requests for them. Instead of returning the measurement uncertainties, the LLM often made them up based on 10% of the measurement value rather than admitting that there were no published uncertainties.

---

[5] https://ned.ipac.caltech.edu/level5/Sept17/Bresolin/frames.html, drawn from "Outskirts of Galaxies," Astrophysics and Space Science Library, Volume 434. ISBN 978-3-319-56569-9. Springer International Publishing AG, 2017, p. 145, https://ui.adsabs.harvard.edu/abs/2017ASSL..434..145B/abstract

More generally, the LLMs made assumptions about the research approach and implemented them without checking first that the student wanted to make those assumptions. The LLMs did not justify the assumptions with references to where similar assumptions had been made. Thus it was hard for the student to judge the merit of the assumptions and to act on their implementation.

**Reflections**

By the semester's end, each student had completed a draft research paper on Overleaf, using the Open Journal of Astrophysics[6] template available there. All papers included the results of novel data analyses or simulations and addressed an astronomy question not yet answered. While no drafts were ready for submission to the journal at that time, students noted that the projects had taught them about new astrophysical topics as well as the advantages and disadvantages of using LLMs for scientific research. The students learned ways that the models could make them more (and less) productive, e.g., by producing a fast, but sometimes inaccurate, synthesis of the relevant literature.

After the completion of the semester, students indicated that they would employ LLMs for research again, but in limited applications and, for one student, only as a last resort. In particular, students would use an LLM 1) to find papers in the literature if they did not know where to start, 2) to fix minor (e.g., syntax) errors or expand on code (e.g., for speed-ups with parallel processing) that they had already written, and 3) in cases where they had experience iterating over LLM outputs, to generate plotting or data analysis scripts after they knew exactly what they wanted.

While recognizing the potential of agents, students would avoid asking current models to generate code to access and download data from online archives. One student was leary of using the models to identify open problems, especially in a field about which the student had limited knowledge. However, another commented that interacting with an LLM made them think more critically about the questions they were asking and how they asked them. The student found that the LLM produced different answers depending on the phrasing of the prompt, but that the back-and-forth interaction taught them how to go from a very broad open question to a problem specific enough to lead to a research project design.

As to whether LLMs saved time in doing research, student responses varied, with about half stating yes. For one student, time was saved "given that my baseline knowledge of modern galaxy research was very little when first formulating project ideas." For another, LLMs were useful for making plots for quick visualization purposes. A third student had grown to use the models more efficiently during the semester and suggested that they might be more helpful in the future. A fourth agreed and found value in now understanding the limitations of LLMs and in what circumstances model use might boost productivity. On the other hand, a fifth student commented that they had learned little directly from the models—knowledge essential for their research project was obtained through their own reading of primary references.

Given their experiences, students provided some ideas for other scientists who want to conduct an original research project with LLMs. Students suggest learning upfront what the models do well and leveraging them for those tasks only. One must be particularly cautious in using LLMs to develop ideas for working on open problems outside a researcher's field of expertise, because it is that much harder to critically examine the outputs. In particular, for a scientist starting a project in an unfamiliar field, LLMs can struggle with providing detailed answers to questions or in-depth insights to open questions in that subject area, especially questions that could be addressed within a several-month timeframe.

While LLMs can be useful in providing references to overview articles and annual reviews, they can flounder in pointing towards more recent papers or those in more niche areas of research. Scientists should be skeptical of citations and links to research papers output by LLMs and always verify that the output summaries align with the studies that they claim to be describing. Manual review of the primary references is key to ensuring accuracy.

---

[6] https://astro.theoj.org/

Asking different LLMs to cross-check each other improves the results by leveraging the different strengths of the models, e.g., a student found that Gemini tended to be more rigorous in its scientific reasoning but lacked creativity, while ChatGPT was more creative but could produce problematic code or results. In addition, the LLMs should be prompted to explain their reasoning before the next prompt. Furthermore, if an LLM keeps generating an incorrect result, re-prompting it without providing additional information does not improve accuracy. It is also helpful to start a new chat, as continuing in the same chat seems to reinforce, not resolve, earlier misunderstandings.

Students advise LLM developers to focus on several areas of improvement. LLMs for scientific applications must be forced to preserve complete citations and data from reputable databases (e.g., ADS) and forbidden from generating new citations or data that do not come directly from such a source. It would be useful if LLMs could clearly communicate their uncertainty or confidence in their responses. Models also must be more nuanced when stating that an idea or project is novel when related concepts have been explored before in the scientific literature. LLMs should be made more agent friendly, i.e., it should be easier to use system prompts to automate different chats with teams of AI reviewers, coders, program managers, etc., and to moderate the interaction among chats with different roles. LLMs should have the capability of generating code or agents that can access niche online datasets, packages, and APIs, e.g., older packages or those with more limited documentation available on platforms like GitHub or ReadtheDocs.

Students generally felt that LLM use detracted from the experience (and fun) of thinking about their research. One student did not like the loss of autonomy when LLMs tried to predict next steps in the research process and was particularly concerned about the impacts on younger researchers with less experience who may just accept each suggested next step without thinking deeply. Students expressed discomfort with any use of LLMs to write the text of research papers; in addition to the ethical implications, current LLMs lack the ability to discuss the science at a publication-quality level of detail and accuracy. Even if models improve to the point of significantly increasing research productivity, students raised concerns about the stifling of creativity and reflection on the research process. One student would use the models "cautiously, as one may lose the 'human-ness' or critical thinking that makes us want to do astrophysics research," e.g., Hogg 2026. Another noted that they would "avoid using LLMs when thinking about the science I want to do, otherwise what's the point of writing a paper?"

Fall 2025 was the first implementation of using LLMs to complete original research over a semester in a graduate course in the University of Arizona Astronomy department. In retrospect, students would have benefited from one class period and/or several early homework problems devoted to understanding LLMs, their limitations, and the best current practices for their use in research. Such experiences could expose students to some of the benefits and pitfalls of LLMs, allowing them to decide to what extent they wish to use LLMs going forward. In future semesters, students will be encouraged early in the term to review relevant and free online resources, e.g., DeepLearning.AI[7], Github[8].

Considering the rapidly improving LLM landscape, some of the problems encountered by students in Fall 2025 and discussed here, such as frequent coding and citation inaccuracies, may be resolved before the next academic cycle. If not, more discussion will be devoted to using LLMs to generate notebooks or agents for the students' scientific workflows. An example workflow might be to prompt LLMs to identify data catalogs or APIs with python access, validate those sources, generate access code for the data with astroquery, iterate on the code, then structure the data into a specific table format and/or produce a plot.

---

[7] https://www.deeplearning.ai/courses/generative-ai-for-everyone/
[8] https://github.com/NASA-IMPACT/LLM-cookbook-for-open-science

While there are many ethical and practical considerations that accompany the heavy use of LLMs, AI is rapidly being adopted by both academic and industry jobs. Among important points for future classroom discussion will be the extensive water and energy requirements for data centers, the environmental impacts of those centers, implicit biases included in training LLMs, corporate conflicts of interest, and lagging government regulation. With these in mind, it is critical that students are trained to succeed in any career they wish to pursue. Students that adapt to use LLMs expertly, creatively, and efficiently will excel.

**Appendix**

Here we list the full comments from each student. Each student is identified only by number, e.g,, [1], [2], …[7] to preserve anonymity.

*What science question were you trying to address? Generally speaking, what kind of measurements and/or calculations were required to address the science question?*

[1] The Impact of Bars on Flattening Metallicity Gradients in Star Forming Galaxies

We had to measure the gas-phase metallicities of galaxies using the O3N2 strong-line calibrator and derived radial metallicity gradients by fitting linear models to the metallicity profiles normalized by effective radius, comparing the slopes of these gradients between barred and unbarred galaxies.

[2] The Impact of Regional Metallicity on the YSO Mass Function in the Local Group (conducted jointly by two students)

Student 1 collected samples of young stellar objects (YSOs) in nearby Milky-Way star forming regions, and in star forming regions of other Local Group galaxies, including the Large and Small Magellanic Clouds (LMC, SMC) and NGC 6822, e.g., from *Spitzer* imaging surveys. Pre-computed stellar mass estimates from the aforementioned surveys were cross-referenced with 12 + log(O/H) measurements that served as proxies for the local gas-phase metallicities surrounding the YSOs to examine the dependence of YSO mass distributions with Local-Group metallicity.

Student 2 collected samples of protostellar cores identified through the Herschel Gould Belt Survey, including the Orion B, Ophiuchus, Taurus, Aquila, and Lupus Molecular Clouds. The stellar masses had already been computed from previous works. These were compared with 12+log(O/H) measurements across the different molecular clouds, which also served as a proxy for the metallicity. We then produced an initial mass function with dependencies on metallicity to probe how metallicity impacts the initial mass function. Because metallicity decreases with redshift, we aimed for metallicity to be a proxy for redshift, probing how the IMF changes with redshift.

[3] Chaotic Evolution of H3-Like Stellar Triples under Dark-Matter Subhalo Perturbations

To address whether dark-matter subhalo flybys can measurably alter the long-term stability/chaotic evolution of very wide hierarchical stellar triples, the project required (1) measuring/setting up the triples' initial orbital architecture—masses and hierarchical orbital elements $(a,e,i,\Omega,\omega)$ and stability metrics; (2) calculating the internal secular evolution (Kozai–Lidov/chaotic secular dynamics including relevant corrections like GR/tides, and/or validating with direct N-body integration); (3) calculating the external perturbations from subhalos, i.e., sampling an encounter population (subhalo mass profile such as NFW, impact parameter, relative velocity, encounter times/rates) and converting each flyby into tidal/impulsive "kicks" with adiabatic suppression; and (4) comparing outcomes (changes in energy/angular momentum, eccentricity/inclination excursions, instability/ejection/collision rates) against a tide-only control to isolate the subhalo-driven effect.

[4] The Influence of Different Stellar Populations on the Integrated UV Spectra of Galaxies

We required a grid of stellar evolution models and their synthetic spectra to generate integrated spectra for galaxies. We needed to make assumptions about the initial mass function of the population, its star formation history, etc. To compare to observations, we required integrated spectra of galaxies.

[5] joint project with [2] above

[6] Investigating Galaxy Bar Evolution via Comparing Theoretical and Observational Data

For this investigation, the student compared numerical simulation results based on two different theoretical models to real data to test which was more consistent with observations. In the first theoretical model, galactic bars evolve within dense halos and experience strong dynamical friction. In the second, bars in maximal disk or low density halos evolve largely through internal disk processes. Measurements of bar strength, bar length, and the corotation ratio were obtained from the literature and compared to the predictions of both models.

[7] Breaking the Slitless Degeneracy: Paschen-a Rotation Curves at z~1 from JWST/NIRCam R&C Grisms

To make kinematic measurements of galaxies at cosmic noon, we used JWST NIRCam grism data in both the R and C directions to calculate the rotation speed of z~2 galaxies using Ha or PaA lines. To achieve this, we needed to develop code for fitting the emission line images. We also needed to identify papers with previous measurements of galaxies at this redshift for cross-validation.

*Had you conducted research on galaxies before this class (Y/N)?*

[1] N
[2] N
[3] Y
[4] N
[5] Y
[6] N
[7] Y

*What was your level of prior experience with LLMs?*

[1] I had moderate experience using LLMs to assist in learning and research.
[2] I have used LLMs to help with some coding for my research and as a supplemental study aid.
[3] I have used LLMs to assist my coding and also let it check what errors I might ignore when the code didn't produce expected results.
[4] I have used LLMs to assist with generating blocks of code that I later iterate over or modify. I primarily use it with generating plotting or data analysis scripts.
[5] I have mostly used LLMs to assist with studying for exams by summarizing my notes and only on rare occasions to assist in making plots related to my research.
[6] I have used LLMs to help fix code and address questions about lectures or research that I had trouble otherwise finding quickly. I have also implemented LLMs to help me study by generating notes, flash cards, or example questions.
[7] I have used LLMs to help write code and summarize papers.

*Which LLM(s) did you use for your project (please also list the version if available)?*

[1] I used ChatGPT5 for the project.
[2] ChatGPT-5 + Claude-4, 4.5
[3] ChatGPT-5.2 + Claude-4+Gemini
[4] I used ChatGPT-5.
[5] ChatGPT-5
[6] ChatGPT-5 and Claude-4
[7] ChatGPT-4o, 5, 5.2

*How did you use them exactly and for what tasks for your project?*

[1] I used LLMs for tasks such as synthesizing relevant literature, brainstorming research questions and methodologies, identifying datasets, and assisting with writing code for analysis. They helped streamline the literature review and early research design, but struggled with generating functional code for data analysis, requiring me to manually correct large portions of the code.

[2] I used the LLMs to help refine our research topic and to identify relevant references and datasets for the project. This sped up the literature review; however, links to the wrong sources were too frequently provided in the process, or papers were misattributed to the wrong citation.

[3] (1) code-level help (explaining what scripts do, rewriting/refactoring Python, and generating reproducible snippets for your astronomy pipelines); and (2) "agentic" / multi-step workflows for research engineering, specifically LLM-driven code-generation loops (e.g., a Gemini-API orchestrator with reviewer/feedback iterations) to help build and debug components of my analysis/simulation tooling.

[4] I used the LLM to get ideas for possible topics to work on and to identify some of the open questions in the field. After deciding on a topic, I used it primarily for finding the exact science question to work on, and the best tools that were available to complete the task. I used STARBURST99 to generate synthetic spectra for stellar populations, and I used the LLM to understand how to use the code and how to extract the specific output that I needed for my project.

[5] I mostly used LLMs to find references and conduct the background research needed for this project. It was somewhat useful in helping to refine the topic ideas, though I mostly came up with the idea for my project without the use of LLMs, because I found that the LLMs struggled to find open questions in a small enough field that I could approach one in the given timeframe. I also used LLMs to help me look for existing models that I could compare my data with.

[6] Having little prior knowledge about galaxy evolution, I used the LLMs to help me figure out a feasible project for someone at my experience level. I asked the LLMs to return information about ongoing research topics and open questions. They found references and summarized them. I also had the LLMs suggest how to access the data I would need, help write code to handle the data, and identify what corrections might be needed to make comparable plots across datasets.

[7] When developing new code, e.g., for fitting suites and figure plotting, I used LLMs to refine and clarify the code structure. Then, I could simply adjust the code based on what I actually wanted. I also tried to use LLMs to locate, translate, and summarize the papers and textbooks when I only wanted a simple conclusion.

*How much time did you spend using them for your project?*

[1] I used the LLM for a few hours during each stage of the research project with some of the time being used to explore what it can and cannot do well.

[2] I spent probably around 3 or 4 hours total using the LLMs.

[3] Used roughly 30 hrs on LLMs. The bottleneck is to reproduce a similar dynamical environment in the MW's stellar halo. Also, it took longer and longer for the LLMs to think as the project grew (from a few seconds thinking to 10's of minutes thinking per prompt).

[4] I probably spent 30min to an hour using LLMs for each stage of the project, about 3 hours in total.

[5] I spent roughly 5–6 hours using the LLMs during the literature search. I eventually grew frustrated with the mistakes they often made during this stage and approached the rest of this background research more similarly to how I would in my other research—without the help of LLMs. I did spend 1–2 hours using LLMs while searching for additional data and models, which they were somewhat more adept at.

[6] I probably spent 5–6 total hours using the LLMs until I ran out of uses. Most of the time was spent going around in circles, telling the LLM that it was doing something wrong and trying to get it to fix itself.

[7] I spent about 10 hours asking LLMs about the related papers, verifying each resource, and also asking them to correct the bugs in my code.

*How did you check the results?*

[1] I checked every source it gave, followed up methodologies it proposed with what is done in the literature, and ensured all the code was doing what I expected it to do.

[2] I had to double-check references with Google and the Astrophysics Data Server (ADS) to ensure that the citations and summaries provided by the LLMs were accurate, which was entirely true in maybe 80% of cases. I read the relevant sections of all the sources identified with the LLMs in order to cite their material directly, not from any generated summary.

[3] I let the LLM explain what it was doing for every few agentic iterations, and I checked through the explanation to see what was wrong. When the LLMs couldn't resolve some issues (e.g., when the relation between code design and the error was too implicit), I'd look into the code it wrote and help fix it. I also let different LLMs cross-check each other.

[4] I tried to read primary references wherever I could and learn about the research area/problem, which helped me verify my results.

[5] Going into this project, I was most concerned about the LLMs making up fake references or referencing the wrong paper for a given result. Because of this, for each reference that the model suggested, I made sure to at a minimum click each link or find the paper on ADS to make sure it existed. If the paper that the LLM referenced seemed like it might be useful to the project I was approaching, I read through the abstract and asked the LLM for the section of the paper that it thought the information had come from. This made it easier to check that the information the LLM gave matched with what was actually in the paper.

[6] I made sure to click each reference the LLM provided and read through the article to check that it was a real paper and actually contained the results cited by the LLM. Whenever the LLM helped to download

data, I used my own written script to check the values, formatting, etc., and made sure that any plot I made seemed reasonable by comparison to others in the literature.

[7] For code modification, running the code immediately told me whether that was what I wanted. For finding and summarizing papers, I went through the abstract and figures quickly to check if there were some obvious errors.

*Which of the tasks did the LLM(s) allow you to complete in an accurate and time-saving way?*

[1] The LLM allowed me to efficiently synthesize relevant literature, identify key papers, and generate research questions and methodologies, significantly speeding up these initial stages of the project. It also helped draft sections of the paper, improving writing efficiency.

[2] Time was saved in my literature search by using the LLMs, but the accuracy was not satisfactory. More often than not, around 80% of the time, it was fine, but the 20% of confused citations and links to the wrong sources did not instill confidence that the results provided were entirely trustworthy. Generating initial project ideas in a broader sense was then likely the best time-saver, and not subject to much inaccuracy.

[3] Only simple code debugging and producing images saved me time. ChatGPT and Gemini can't produce more complex jobs such as simulation code design faster than me. Claude-4 is the only LLM that can do a more complicated job, but the usage limit is also lower than ChatGPT and Gemini.

[4] I think the LLM allowed me to narrow down the specific goal of the project and the tools that I could use to answer the research question.

[5] LLMs did help me to conduct my literature search and narrow down the questions I was interested in by pointing me towards papers that addressed some of the major open questions in galaxies.

[6] LLMs helped me most through finding a topic to work on, answering questions regarding galaxy evolution that were not answered in my class notes, and summarizing past research.

[7] LLMs helped me to write code very quickly, which allowed me to do the technical part faster and saved my time for thinking about the science.

*In your interaction with the LLM(s), were there any positive surprises?*

[1] A positive surprise was how effectively the LLM synthesized complex information from various sources and quickly provided summaries and insights that guided the research direction. Since this is not my area of research, it would otherwise have taken many days or weeks to read through papers to inform my decision for a novel research topic.

[2] I was surprised that the LLMs seemed to generally be pretty well trained on astronomy and astrophysics literature, if sometimes still confusing sources. I was not a topic expert on galaxy research, and the LLMs had a wider and deeper knowledge of the field overall than I did when beginning the project.

[3] Not much surprise with ChatGPT and Gemini for coding. But ChatGPT surprised me that it could point out what was wrong in the code by looking at the energy and angular momentum evolution plot. Claude-4 is the most impressive LLM in that it actually solved the complicated problem that ChatGPT and Gemini couldn't solve.

[4] I think the LLMs have a reasonable amount of "knowledge" about various fields and research questions, and can answer basic questions about them. I could see that as a positive point.

[5] Because I hadn't used LLMs very much for things other than summarizing my notes and readings prior to this class, I was surprised that it would suggest potential next steps in what I was asking it to do. I was also surprised that it was reasonably adept at writing code, especially for small and specific tasks.

[6] I was interested in how one LLM could view the plots/images in a paper without issues and another might have difficulty just opening the paper or viewing text.

[7] The LLMs know physics better than I thought. I worried about LLMs making many scientific errors, but the more common situation was that they taught me a lot about physics when I could not understand the paper myself.

*Which of the tasks did the LLM(s) fail to help you complete?*

[1] The LLM struggled with generating functional code for data analysis, particularly when working with the MaNGA dataset, as the code it produced often needed manual correction in order to run properly. It also faced limitations in accessing and retrieving observational datasets autonomously.

[2] The LLMs were not capable of retrieving and downloading survey data directly from archives like VizieR. While they may often be able to point one to the correct link to the data tables, they did not seem to have the ability to query the tables themselves. This is one reason why I found the LLMs to be most useful just for the initial literature search.

[3] When the LLMs and I attempted to introduce a flyby subhalo into the simulation, we frequently encountered non-physical results, and the LLMs were unable to diagnose the issue. The problem was ultimately traced to an incorrect implementation of the Milky Way's tidal field, as well as an incompatible formatting of the subhalo density profile within the simulation package. Correcting the tidal setup and implementing the subhalo profile in a package-compatible format resolved the issue.

[4] The LLMs failed to be of much help in using the specific tool (STARBURST99) for my project, and I had to figure out most of it myself.

[5] One of the major tasks I asked the LLM to help me with was determining which databases the archival data I wanted to use were stored on and how to access them. The LLM consistently identified the wrong database and confidently wrote code to access data from archives that do not exist. When I was searching for data about the metallicities of each of the molecular cloud regions, the LLM also often provided values without references or errorbars, even when I repeatedly requested them. Particularly with errorbars, the LLMs often made them up based on 10% of the value, rather than just telling me that the values did not exist in the literature.

[6] The LLM failed to access the correct data I needed and to format it similarly to the comparison samples. Often the dataset returned had size index errors or was empty.

[7] The LLMs could not write good code when a package was less common or was updated very frequently. For example, when I asked the LLMs to use photutils, they always returned code using the discontinued photutils 1 instead of the more commonly-used photutils 2. After being asked for a paper summary, the LLMs sometimes returned the wrong numbers and conclusions. When asked to find papers, they usually ended up with the wrong summaries and citations, even when the links were correctly returned and the authorship was correctly attributed.

*Did the failures surprise you or not and why?*

[1] The failures did not surprise me because I expected that LLM would struggle with tasks requiring code specific to a database such as MaNGA.

[2] I was surprised that the LLMs could not more readily assist in retrieving data from online archives. However, maybe it should not be so surprising since this may be a niche task relative to the needs of typical chatbot users.

[3] No for both cases. The tidal effect is implicit to the error, and the simulation package I used is not well documented, so it's natural that the LLMs couldn't fix these issues.

[4] Most of them did not surprise me, as I had experienced them before. I guess I was a little bit surprised by the LLMs doubling down on something that was clearly pointed out to be wrong.

[5] It surprised me that the LLMs seemed to refuse to respond that the answer to my question did not exist in the literature that it could access. In particular, it surprised me that the LLMs could write code to find data that did not exist. It also surprised me that the LLMs would make assumptions about my research approach and implement them without checking with me that these were assumptions I wanted to make. These assumptions also were never justified with a reference to where similar assumptions were previously made, so they were not ones that I was able to easily implement.

[6] Knowing how to prompt LLMs is a skill in itself, so I was not surprised when the results varied or failed depending on how I asked.

[7] Even if I provided the LLMs with a paper or link and asked them to return a data table in the paper in a machine-readable way, they sometimes returned numbers totally different from those in the provided source.

*Were there certain failure modes that the failures had in common, e.g., problems writing code to access external APIs, returning inaccurate references or links, returning code that would not compile?*

[1] The failures were primarily related to generating code that was incompatible with the MaNGA dataset format.

[2] I think that the main weakness I encountered was an apparent inability to directly search the internet or to directly interact with APIs, such as for querying data tables and downloading data from them. The LLMs were also not necessarily familiar with the technical details of such datasets and how they are organized internally beyond being able to point to their URLs.

[3] Returns a code that doesn't produce physically reasonable results

[4] It returned inaccurate references, gave very generalized broad answers for specific questions that weren't useful, and doubled down on things that I had clearly shown were wrong—for example, it kept pointing towards lines in a piece of code that did not exist.

[5] It returned some code that would compile, but it would make assumptions or overly simplify the problem that I was asking. As an example, if I asked for different model initial mass functions to all be plotted on the same graph with references, the LLM would plot a single and basic IMF without any references. It would explain that this fit the general shape of an initial mass function, but did not address the purpose of the question I asked.

[6] Neither of the LLMs I used performed well in accessing data or writing functioning code.

[7] The LLMs can fail when asked questions that are open or beyond the knowledge of the LLMs. In addition, the models have trouble returning data accurately, e.g., measurements and their uncertainties.

*Would you use LLMs for your research again?*

[1] Y
[2] Y, in some limited applications.
[3] Y
[4] Y, for limited purposes.
[5] Y, though as a last resort and in limited applications.
[6] Y
[7] Y

*If so, in what ways?*

[1] I would use it to find literature if I did not know where to start. I would also use it to find minor errors or expand on code I have already written.

[2] While not a core use of LLMs for this project, I have otherwise found them to be useful for tweaking existing codes (e.g., for speed-ups with parallel processing). In my own research, I do not rely on LLMs for literature searches, as I have a better understanding of my own field and its researchers, which allows me to directly search ADS efficiently.

[3] For simpler cases such as plotting and fixing syntax errors. More complicated cases require agentic use of LLM (i.e., making different LLMs play different roles such as reviewer, debugger, code writer, etc.).

[4] I would use it for generating coding/plotting/data analysis scripts for my research. I feel comfortable doing so, because I have enough experience with coding to iterate over the LLM output, and would overall save some time on laborious work.

[5] I would consider using LLMs to help me with specific aspects of code for plots in my research. LLMs can be especially useful for designing plots to be more elegant when I have a specific vision for how I want them to look. I think the models can also be useful in understanding how to use certain new packages, although the models can struggle with interpreting other packages.

[6] I would use it to help me debug code that I already wrote or for when I have to quickly make a plot.

[7] I would use the models for writing code and then modify the code myself.

*What kinds of things did using the LLMs teach you?*

[1] It taught me how I can use it as a tool to make myself more productive, and the areas where it makes me less productive.

[2] Using the LLMs taught me to be more cautious about the citations that they output, which can sound or look like they could be correct, or may even be just slightly off from the truth. It became clearer to me where the limits of LLM abilities were by the end of the project.

[3] You need to understand what the LLM is doing before using LLMs to further improve the code, otherwise when there's an error that LLMs cannot solve you won't be able to fix it. For example, when I included the Galactic tidal force, the LLM cited two papers describing the Milky Way's potential and combined them. Initially, I was only aware of one of the references and found that the code was producing an unreasonable potential. Upon closer inspection, I realized that it had mixed parameters from two different papers. After replacing the parameters to make them consistent, the results became reasonable.

[4] It taught me about some of the very broad research areas and questions related to galaxies, which is a field that I am not familiar with. However, I think this could have been achieved through other means as well, such as reading reviews, textbooks etc.

[5] The LLMs taught me to think more critically about the questions I am asking and how I ask them. It could give very different answers depending on how I phrased things. As I attempted to use the LLMs to determine a specific question to research, I also learned how to go from a very broad open question that the LLM could provide to a more specific question that I could design a research project around.

[6] Using LLMs taught me that learning how to phrase your questions is very important and that knowing how to correct an LLM requires some knowledge in the field/topic you are asking it about in the first place.

[7] Using LLMs significantly shortened my time dealing with simple, formal tasks that did not require much thinking, e.g., writing plotting code. They were poorer at helping me with more complex science-related tasks, e.g., coming up with a way to measure the R and C grism data at the same time.

*For what aspects of your research would you avoid using LLMs?*

[1] I would avoid asking it to generate code to access data in a dataset.

[2] I would avoid using LLMs to write the text of research papers or to download data from online archives. They should only be used very carefully to identify references/citations.

[3] I'd avoid using LLMs when thinking about the science I want to do, otherwise what's the point of writing a paper?

[4] I would avoid asking it to come up with open problems to work on, especially in a field that I have limited knowledge about. Maybe at some point they will improve and increase productivity in research, but even then, I would use them cautiously, as one may lose the "human-ness" or critical thinking that makes us want to do astrophysics research. I agree with many of the points put out in this paper: https://arxiv.org/abs/2602.10181v1.

[5] I would avoid using LLMs to help me find a narrow enough research question to approach in a single research project. I would also avoid having LLMs write any text that I plan on using in a paper or proposal because they lack the ability to discuss the science in enough detail. I would also avoid having LLMs write entire pieces of code. While the models can be useful for small pieces or lines where the notes are buried in the documentation, asking a LLM to write the code for the entire analysis usually results in false assumptions or important steps being skipped.

[6] I would avoid using it to write the text of papers or to make a quick list of references.

[7] I would avoid using LLMs when they do not solve a specific coding issue after a few tries. For example, if a plotting layout is consistently incorrect, re-prompting without additional information does not

help. Continuing in the same chat seems also to reinforce, not clear up, earlier misunderstandings. In such cases, I prefer to stop and debug the problem myself.

*On balance, did the LLMs waste or save you time?*

[1] I think overall it saved time in the research process.

[2] Overall, I saved time by using the LLMs, especially given that my baseline knowledge of modern galaxy research was very little when first formulating project ideas. Performing a thorough literature review from scratch without the aid of LLMs would have extended the time for me to get up and running on the project.

[3] Wasted my time, but now I learned more about how to work with LLMs so I will work with them in a more efficient way next time. Hope that saves me time in the future use of LLMs.

[4] I think overall I did not save any time using LLMs and did not learn much directly from them. Most of what I learned about the specific research area was through my own reading of primary references.

[5] I don't think they saved me very much time, but I don't think they were a complete waste of my time either. I did learn some ways to better ask LLMs questions and what skills they are better at, so that if I want to use LLMs more in my future research, I know what their limitations are. I think this experience of using LLMs will ultimately save me time because I know how to use them better and in what circumstances attempting to use LLMs will save or waste my time.

[6] For this course specifically I think using LLMs helped me to develop some knowledge and produce results for the project quickly, even though the results were not perfect or always correct.

[7] I think LLMs saved me time in certain cases, e.g., making plots for quick visualization purposes.

*Given your experience working on an original research project with LLMs, what would you suggest to other scientists who want to do so?*

[1] Identify what it can do well and leverage it for that, but do not waste time trying to get it to do things it seems to be struggling with.

[2] I would suggest that scientists be very cautious of citations and links to research papers that are output by LLMs, always verifying that the summaries provided by the chatbot(s) align with the studies that they claim to be describing. The models would be best used for quickly generating ideas in an unfamiliar subject area and/or for generating and iterating upon codes.

[3] I think it's good to brainstorm with LLMs, just make sure the sources the LLMs cite are real and reasonable. Also, it's good to ask LLMs to summarize what they are doing before you give them the next prompt. It's also good to ask different LLMs to cross check each other.

[4] I would suggest not using LLMs to develop ideas for working on open problems in a field that one is not familiar with. I would only be comfortable using them if I have some reasonable amount of knowledge about the field that can help me critically examine the LLM output. Perhaps they could be used to ask for references where one could learn about a new research area, and then use the primary reference to do the actual literature review.

[5] I would caution other scientists from using LLMs. I think when you first start working with them, they can come across as magical or as a new power. But if you dig deeper than a surface level, especially in an area that you are much more familiar with, I think the appeal can drop. I think when using LLMs, it is essential to assume more of what they respond is incorrect until you can verify that it is correct and from a source you trust. If you aren't willing to put in this additional work some significant fraction of the time, then using a LLM in your research may not be the best approach for your research.

[6] Make sure to verify any information that LLMs return with outside sources like textbooks, other scientists, etc.

[7] Check very carefully the measurements the models return, even when a source for the data is explicitly provided.

*What would you suggest that LLM developers improve?*

[1] I would suggest attempting to make it have the ability to generate code that knows how to access niche datasets.

[2] I think that LLMs for scientific applications need to be forced to preserve complete citations from reputable databases (e.g., ADS), and they should be forbidden from generating new references or citations which are not directly coming from such a source. It would also be useful if LLMs could clearly communicate their uncertainty or confidence in their responses.

[3] Make it more agent friendly (i.e., create a project that already gives different chats with system prompts for reviewer, coder, program manager, etc.) and automate the interaction between chats with different roles.

[4] I found the LLMs to generate fake citations and say that something is novel and never done before even when it has. Perhaps the "truthfullness" of LLMs could be improved.

[5] There are a few things that I would suggest that LLM developers improve. Several of my peers have already suggested that LLMs should be unable to generate fake citations and that there should be an additional step in place for the LLM to check that the references are correct before they are suggested. I also agree that LLMs could be made more useful if they were better able to learn to write code in niche languages or use niche packages, particularly older packages or those with more limited public documentation on platforms like GitHub or ReadtheDocs. But the thing that I would most like to see in LLMs in the future is for the suggestions at the end of a response to be minimized. I did not like the way that the LLM would try to predict my next steps in the research process. I thought it took away some of my autonomy as a researcher. I worry how it would impact even younger researchers with less research experience, who may just accept each suggested next step in the research process without thinking about what the questions they wanted to answer and how they might answer them without potential answers being spoon-fed to them. This seemed to stifle my creativity and reflection on the research process.

[6] I would hope developers could improve how LLMs verify their results, such as to prevent returning fake citations, etc.

[7] The developers may need to create the means, e.g., through agents, to ensure returned data and citation accuracy. Because LLMs can access the internet and find paper URLs, the paper summaries and citations should be forced to come directly from the source.